\documentclass[fleqn,usenatbib]{mnras}

\usepackage[T1]{fontenc}
\usepackage{ae,aecompl}
\usepackage{lineno}
\usepackage{graphicx}	% Including figure files
\usepackage{amsmath}	% Advanced maths commands
\usepackage{amssymb}	% Extra maths symbols
\usepackage{threeparttable}
\usepackage{multirow}
\usepackage{url}
\usepackage{multicol}
\usepackage{xcolor}

\newcommand{\HI}{\mbox{\rm H{\small I}}}
		
\newcommand{\mum}{\ifmmode{\rm \mu m}\else{$\mu$m}\fi}

\newcommand{\chisq}{\ifmmode{\chi^{2} }\else{$\chi^2$}\fi}
\newcommand{\rchisq}{\ifmmode{\chi^{2} }\else{$\chi^2_\nu$}\fi}

\newcommand{\HII}{H\,{\sc ii}}

\newcommand{\jhk}{\emph{JHK}$_{s}$}

\usepackage{orcidlink}
\usepackage{academicons}
\definecolor{orcidlogocol}{HTML}{A6CE39}

\usepackage{newtxtext,newtxmath}
\title[{\em JWST} NGC 6822: Stellar Populations]{JWST MIRI and NIRCam Unveil Previously Unseen Infrared Stellar Populations in NGC 6822}

\author[Nally et al.]{Conor Nally$^{1}$\thanks{E-mail: conor.nally@ed.ac.uk}\orcidlink{0000-0002-7512-1662},
Olivia C.\ Jones$^{2}$\orcidlink{0000-0003-4870-5547},
Laura Lenki\'{c}$^{3,4}$\orcidlink{0000-0003-4023-8657},
Nolan Habel$^{4}$\orcidlink{0000-0002-2667-1676},
\newauthor{Alec S.\ Hirschauer$^{5}$\orcidlink{0000-0002-2954-8622},
Margaret Meixner$^{4}$\orcidlink{0000-0002-0522-3743},
P.\ J.\ Kavanagh$^{6}$\orcidlink{0000-0001-6872-2358},
Martha L.\ Boyer$^{5}$\orcidlink{0000-0003-4850-9589},
}
\newauthor{Annette M. N. Ferguson$^{1}$\orcidlink{0000-0001-7934-1278},
B.\ A.\ Sargent$^{5,7}$\orcidlink{0000-0001-9855-8261},
Omnarayani Nayak$^{5}$\orcidlink{0000-0001-6576-6339} and 
Tea Temim$^{8}$\orcidlink{0000-0001-7380-3144}}\\
\\
$^{1}$ Institute for Astronomy, University of Edinburgh, Blackford Hill, Edinburgh, EH9 3HJ, UK \\
$^{2}$ UK Astronomy Technology Centre, Royal Observatory, Blackford Hill, Edinburgh, EH9 3HJ, UK \\
$^{3}$ Stratospheric Observatory for Infrared Astronomy, NASA Ames Research Center, Mail Stop 204-14, Moffett Field, CA 94035, USA \\
$^{4}$ Jet Propulsion Laboratory, California Institute of Technology, 4800 Oak Grove Dr., Pasadena, CA 91109, USA \\
$^{5}$ Space Telescope Science Institute, 3700 San Martin Drive, Baltimore, MD 21218, USA \\
$^{6}$ Department of Experimental Physics, Maynooth University, Maynooth, Co. Kildare, Ireland \\
$^{7}$ Department of Physics \& Astronomy, Johns Hopkins University, 3400 N.\ Charles St., Baltimore, MD 21218, USA \\
$^{8}$ Department of Astrophysical Sciences, Princeton University, Princeton, NJ 08544, USA
}
\date{Accepted XXX. Received YYY; in original form ZZZ}

\pubyear{2024}

\begin{document}
\label{firstpage}
\pagerange{\pageref{firstpage}--\pageref{lastpage}}
\maketitle

%--------------------------------------------------------------------------------------------------------------

\begin{abstract}
NGC~6822 is a nearby ($\sim$490~kpc) non-interacting low-metallicity (0.2~$Z_{\odot}$) dwarf galaxy which hosts several prominent \HII{} regions, including sites of highly embedded active star formation. In this work, we present an imaging survey of NGC~6822 conducted with the NIRCam and MIRI instruments onboard JWST. We describe the data reduction, source extraction, and stellar population identifications from combined near- and mid-infrared (IR) photometry. Our NIRCam observations reach seven magnitudes deeper than previous \jhk{} surveys of this galaxy, which were sensitive to just below the tip of the red giant branch (TRGB). These JWST observations thus reveal for the first time in the near-IR the red clump stellar population and extend nearly three magnitudes deeper. In the mid-IR, we observe roughly two magnitudes below the TRGB with the MIRI F770W and F1000W filters. With these improvements in sensitivity, we produce a catalogue of $\sim$900,000 point sources over an area of $\sim 6.0 \times 4.3$~arcmin$^{2}$. We present several NIRCam and MIRI colour-magnitude diagrams and discuss which colour combinations provide useful separations of various stellar populations to aid in future JWST observation planning. Finally, we find populations of carbon- and oxygen-rich asymptotic giant branch stars which will assist in improving our understanding of dust production in low-metallicity, early Universe analogue galaxies.
\end{abstract}

%--------------------------------------------------------------------------------------------------------------

\begin{keywords}
galaxies: dwarf -- galaxies: irregular -- galaxies: individual (NGC 6822) -- infrared: galaxies -- infrared: stars -- stars: AGB and post-AGB
\end{keywords}

%%%%%%%%%%%%%%%%%%%%%%%%%%%%%%%%%%%%%%%%%%%%%%%%%%

%%%%%%%%%%%%%%%%% BODY OF PAPER %%%%%%%%%%%%%%%%%%

%-------------------------------------------------------------------
\section{Introduction} % Section 1.
\label{sec:intro}

The dwarf irregular galaxy NGC~6822 is one of the closest ($\mathrm{d} \sim 490 \pm 40$~kpc;~\citealp{bib:Sibbons2012, Fusco2012}) in the Local Group. 
It is famously home to some of the largest, brightest star-forming regions in the local universe \citep{bib:Kennicutt1979, bib:Hodge1989, bib:Odell1999, bib:Jones2019} with active star formation throughout its disk and central bar. Its low metallicity ([Fe/H] $\approx -1.2$; $\sim$30\%~$Z_{\odot}$; \citealp{bib:Skillman1989, bib:Lee2006}), elevated star formation rate, and overall youth make it a nearby analogue of galaxies at the universal epoch of peak star formation ($z \sim 1.5-2$; \citealp{bib:Madau2014}), at which point a majority of the universe's star formation and chemical enrichment is expected to have taken place. We note that while NGC~6822 is generally considered to be isolated \citep{bib:deBlok2000,Battinelli2006} without any detectable satellites, \citet{Zhang2021} postulate a 200~kpc close passage within the virial radius of the Milky Way $3-4$~Gyr ago. \autoref{tab:characteristics} lists properties of NGC~6822 that have been adopted for this study.

As a target of many stellar population studies, NGC~6822 has been extensively surveyed in both the optical and infrared (IR). An old ($\sim$11~Gyr) stellar component within the galaxy was inferred with the first discovery of RR-Lyrae stars \citep{Clementini2003, Baldacci2003}, while several localised bright \HII{} regions in the central bar \citep{Cannon2006} show the galaxy is still actively forming stars \citep{bib:Jones2019, bib:Kinson2021}. After steady star formation over its evolution, NGC~6822 began a burst of star formation $\sim$3~Gyr ago \citep{Tolstoy2001} and has continued to increase over the last $100-200$~Myr \citep{Gallart1996c}. Optical surveys with the \emph{Hubble Space Telescope} \citep[HST;][]{Wyder2001} and \emph{Subaru} \citep{Tantalo_2022} detect old and intermediate-age red clump (RC) and red giant branch (RGB) stars in NGC 6822. The asymptotic giant branch (AGB) stars within the galaxy trace intermediate age stellar populations and have been extensively studied in the IR \citep[see e.g.,][]{bib:Cioni2005,bib:Sibbons2012,bib:Hirschauer2020,Tantalo_2022} where they can dominate the IR integrated light of a stellar population \citep{Maraston2005}. These cool luminous stars evolve from low- to intermediate-mass ($\sim 0.8-8$~$M_{\odot}$) main sequence (MS) progenitors on timescales ranging from $\sim$100 Myr to more than 10 Gyr \citep{Bruzual2003, Marigo2010} and (together with supernovae) may play a vital role in the metal and dust enrichment of the interstellar medium and ultimately the chemical enrichment of galaxies \citep[e.g.,][]{Knapp1993, Karakas2009, Matsuura2009, Schneider2014, Salaris2014, bib:Srinivasan2016}. %dust evolution of galaxies.
In NGC 6822 the old stellar population is spherically distributed while the young population has a well-defined bar and a disk-like distribution \citep{Tantalo_2022}.

The dust contribution from evolved stars in metal-poor environments and at high redshift is not currently well defined.
In particular, for young galaxies inhabiting the early universe, it can be expected that insufficient time has elapsed for MS stars to reach the AGB phase, which suggests that they do not contribute significantly to the overall dust budget.
Observations have shown that large reservoirs of dust appear to exist at redshifts out to $z$ $\sim$ 6 \citep{bib:Bertoldi2003, bib:Robson2004, bib:Beelen2006, bib:Algera2023}.
Studies of nearby metal-poor systems such as dwarf galaxies and globular clusters have shown that dust can originate from AGB stars in these environments \citep{bib:McDonald2010, bib:Whitelock2018, bib:Jones2018}, however the effects of metallicity on AGB star dust production are disputed \citep{bib:vanLoon2005, bib:McDonald2011, bib:Sloan2012, bib:Sloan2016}. The Surveying the Agents of a Galaxy's Evolution (SAGE) surveys \citep{bib:Meixner2006,bib:Gordon2011} mapped the Magellanic Clouds with \emph{Spitzer} from 3.6 to 160~\micron{}. These data enabled detailed studies characterising the evolved stellar populations within the Magellanic Clouds and their contributions to dust production \citep{bib:Blum2006,Srinivasan2009,bib:Boyer2011, Jones2015b}. \citet{bib:Boyer2011} find that the very dusty ``extreme'' carbon-rich AGB stars dominate the return of dust into the interstellar medium (ISM) in both galaxies \citep[$\sim$90\% of the dust input from evolved stars;][]{bib:Boyer2012}, while red supergiant (RSG) stars do not contribute significantly ($<$4\%). 

While studies such as SAGE have made great progress towards understanding dust production mechanisms at low metallicity, it is now possible to extend this work to more distant systems with JWST. The JWST imaging program of NGC~6822 (program ID: 1234, PI: M.\ Meixner) that we present in this work allows us to investigate the evolved star population out to $\sim$500~kpc at the resolution and depth that was achieved by \emph{Spitzer} in the Magellanic Clouds. This work builds upon the studies of \citet{bib:Jones2017}, \citet{bib:Hirschauer2020}, and \citet{bib:Kinson2021} that characterised the IR stellar populations of NGC~6822.

This paper presents the overview of the NGC~6822 JWST imaging program, including the observation details, data reduction, photometric extraction, and combined NIRCam and MIRI colour-magnitude diagrams (CMDs) used to identify its various stellar populations, with a focus on the evolved stars.  
Section~\ref{sec:obs} presents the observational strategy in detail. We describe the data reduction and photometric extractions in Section~\ref{sec:photometry}. In Section~\ref{sec:prelimresults}, we present and discuss our resulting images, stellar classifications, and luminosity functions. Finally, we summarise the results in Section~\ref{sec:summary}.

%%%%%%%%%%%%%%%%%%%%%%%%%%%%%%%%%%%%%%%%%%%%%%%%%%
%% SECTION 2: OBSERVING PROGRAM
%%%%%%%%%%%%%%%%%%%%%%%%%%%%%%%%%%%%%%%%%%%%%%%%%%

\section{Observation Strategy} % Section 2.
\label{sec:obs}

\begin{figure*} % Figure 1
    \centering
    \includegraphics[trim=0cm 0cm 0cm 0cm, clip=true, width=0.9\textwidth]{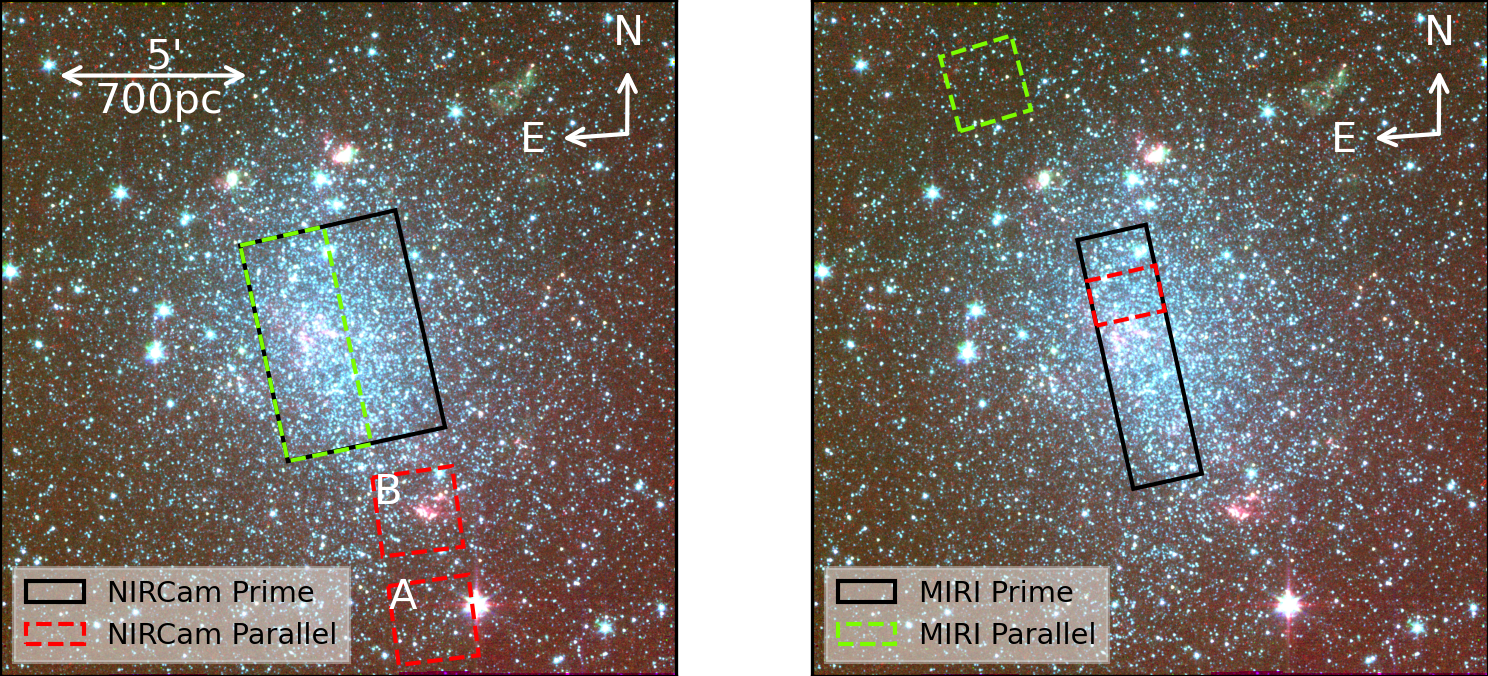}
    \caption{{\em Spitzer} three-colour image of NGC ~6822 from \citet{Cannon2006} with the JWST NIRCam and MIRI survey regions superimposed (solid lines). The left image shows the NIRCam coverage and the right image the MIRI coverage. The NIRCam tile with an associated MIRI parallel is shown in green dashes, while the MIRI tile with its associated NIRCam parallel is in red dashes with detector modules A and B labelled accordingly. North is up and east is to the left.}
    \label{fig:FoVs}
\end{figure*}

We have imaged the central bar of the galaxy NGC~6822 with JWST (Program ID:\ 1234; PI:\ M.\ Meixner), utilizing both the Near Infrared Camera (NIRCam; \citealp{bib:Rieke2005,bib:Rieke2023}) and Mid-Infrared Instrument (MIRI; \citealp{bib:Rieke2015, Wright2023}).
In total, 19.92 hours of integration time were split between NIRCam ($\sim$4.80 hours) and MIRI ($\sim$15.12 hours). 
The spatial coverage for the two instruments, which focus on the central stellar bar of NGC~6822, is illustrated in ~\autoref{fig:FoVs}.
The FoV of both instruments includes the newly-discovered young, massive, embedded super star cluster (SSC) candidate region Spitzer I (e.g., \citealp{bib:Jones2019, bib:Hirschauer2020, lenkic2023}), in addition to a plethora of IR-bright, dusty sources in the galaxy's central stellar bar \citep{bib:Sibbons2012, bib:Jones2019, bib:Hirschauer2020, bib:Kinson2022}.
Additionally, we obtained coordinated parallel observations of off-target regions.

\begin{table} % Table 1
	{\centering
    \caption{Summary of NGC 6822 properties and survey details.}
	\begin{tabular}{lc} 
        \hline
        \hline
        \multicolumn{2}{c}{Properties of NGC 6822}\\
        \hline
        Nominal centre point & 19:44:56 -14:47:51\\
        Distance & 490 $\pm$ 40 kpc \citep{Fusco2012}\\
        Distance modulus $(m-M)_0$ & 23.45\\
        E(B-V) & 0.35 $\pm$ 0.04 \citep{Tantalo_2022} \\
        Metallicity & [Fe/H]$=-1.2$~\citep{bib:Lee2006}\\
        \hline
	\end{tabular}
	\label{tab:characteristics}
    }
\end{table}

\begin{table} % Table 2
	{\centering
  	\caption{Summary of NIRCam observation properties. Note that the quoted values for the PSF full widths at half maximum are that of the empirical PSF.}
	\begin{tabular}{lc} 
        \hline
        \hline
        \multicolumn{2}{c}{NIRCam Observation Properties}\\
        \hline
        Survey area ($arcmin^2$) & 29.0\\
        Total time (hr)& 4.80\\
        Filters & F115W, F200W, F356W, F444W \\
        Central $\lambda$ ($\mu m$)& 1.154,  1.990, 3.563, 4.421\\
        FWHM at $\lambda$ (pixel) &  1.290, 2.129, 1.841, 2.302\\
        Completeness limits at $\lambda$ (mag) & 25.60 24.58 23.21 23.09\\
        \hline
	\end{tabular}
	\label{tab:nircam_params}
    }
\end{table}

\begin{table} % Table 3
	{\centering
  	\caption{Summary of MIRI observation properties.}
	\begin{tabular}{lc} 
        \hline
        \hline
        \multicolumn{2}{c}{MIRI Observation Properties}\\
        \hline
        Survey area ($arcmin^2$) & 14.5\\
        Total time (hr)& 15.12\\
        Filters & F770W, F1000W, F1500W, F2100W \\
        Central $\lambda$ ($\mu m$)& 7.528, 9.883, 14.932, 20.563\\
        FWHM at $\lambda$ (pixel) &  2.445, 2.982, 4.436, 6.127 \\
        Completeness limits at $\lambda$ (mag) & 20.43 19.41 16.80 16.86 \\
        \hline
	\end{tabular}
	\label{tab:miri_params}
    }
\end{table}

\subsection{NIRCam} % Section 2.1.
\label{sec:NIRCam_strat}

Our observations with NIRCam comprised a total of $\sim$4.80 hours of integration time and were obtained on 2022 September 4 utilizing the F115W and F200W short-wavelength filters and the F356W and F444W long-wavelength filters.
The observations were taken with a \textsc{full} subarray and implemented the \textsc{fullbox} \textsc{4tight} primary dither pattern, which ensures that the gaps between the four short-wavelength detector subarrays in each module are filled, as is the $\sim$43\arcsec\ gap between the A and B modules. We positioned the east and west portions of our mosaic such that there is a 10.0\% overlap in both the rows and columns of the resultant 2$\times$1 mosaic.
The \textsc{bright2} readout pattern was selected to optimize the signal-to-noise ratio (S/N), with one integration per exposure and seven groups per integration.
With a three-point sub-pixel dither pattern to sample the point spread function (PSF), this resulted in 12 total dithers, with an overall total exposure time of 1803.8 seconds for each of the short- + long-wavelength filter combinations (i.e., F115W + F356W and F200W + F444W). See also \autoref{tab:nircam_params} for filter properties and appendix \autoref{tab:obs_params_NIRCam} for a complete summary of exposure parameters.

Coverage of NGC 6822's central stellar bar was achieved by restricting the aperture position angle (PA) of the JWST spacecraft to between 92.0$^{\circ}$ and 93.0$^{\circ}$.
Finally, observations of the east and west halves of the mosaic were grouped to be non-interruptible, ensuring a consistent relative orientation. The NIRCam mosaic was centred at RA = 19:44:56.1990, Dec.\ = --14:47:51.29; the tile with a MIRI coordinated parallel is centred at RA = 19:45:00.2644, Dec.\ = --14:47:55.23.

Our selection of NIRCam filters was based on prior work on stellar populations at similar wavelengths.
The F115W and F200W filters are the closest equivalent to standard Johnson \emph{J} and \emph{K$_{\rm s}$} filters, respectively, while the F356W and F444W filters are similar to \emph{Spitzer} IRAC [3.6] and [4.5], respectively.
With these filter selections, we can construct a standard collection of diagnostic near-IR CMDs, such as F115W -- F200W vs.\ F200W (similar to \emph{J} -- \emph{K$_s$} vs.\ \emph{$K_s$}).
This allows for comparative study with previous observations made of other galaxies, including the SAGE studies of the Magellanic Clouds (e.g., \citealp{bib:Blum2006, bib:Meixner2006, bib:Whitney2008, bib:Gordon2011}), the DUST in Nearby Galaxies with Spitzer (DUSTiNGS) project which surveyed local group dwarf galaxies in the mid-IR (e.g., \citealp{bib:Boyer2015a, bib:McQuinn2017, bib:Goldman2019}), and others (e.g., \citealp{Jones2015, bib:Jones2018, bib:Jones2019, bib:Hirschauer2020}).

\subsection{MIRI} % Section 2.2.
\label{sec:MIRI_strat}

MIRI observations totalling $\sim$15.12 hours were obtained between 2022 September 4 and 2022 September 15. The 1$\times$6 mosaic was restricted to a PA of 93$^{\circ}$ and positioned to run along the central stellar bar of NGC~6822. Each pointing in the mosaic included a 10.0\% overlap between the rows to ensure smooth background matching between the fields. Due to an error in the program implementation, only five of the six MIRI tiles in the mosaic were observed on 2022 September 4. The final MIRI tile (and its coordinated NIRCam parallel observations) were subsequently obtained on 15 September 2022. Our program employed the F770W, F1000W, F1500W, and F2100W filters using a \textsc{small} four-point \textsc{cycling} dither pattern, the \textsc{full} subarray and \textsc{fastr1} readout pattern. A list of filter properties can be found in \autoref{tab:miri_params} and a list of exposure parameters including groups per integration and integrations per exposure for each filter is provided in appendix \autoref{tab:obs_params_MIRI}.
The MIRI mosaic was centred at RA = 19:44:58.0949, Dec.\ = --14:48.20.62; the MIRI tile with a NIRCam coordinated parallel was centred at RA = 19:44:58.4923, Dec.\ = --14:46:40.85.

Our selection of MIRI filters was guided by the predicted MIRI colours of mid-IR bright stellar populations from \citet{bib:Jones2017}:
CMDs constructed from these wavelengths reveal the reddest and dustiest sources, distinguishing evolved (RSGs and AGB stars) and young (YSOs) stellar populations.
In addition, the F770W filter traces polycyclic aromatic hydrocarbon (PAH) emission which is expected to be prevalent in star-forming regions, while the F1000W filter is sensitive to the 10~$\mu$m silicate absorption present in the spectral signatures of dusty, embedded YSOs.

\subsection{Parallel Imaging} % Section 2.3.
\label{sec:parallels_strat}

Accompanying our primary pointings, we obtained parallel observations in both NIRCam and MIRI. Only one of the two 2$\times$1 mosaic tiles in our primary NIRCam observations has an associated MIRI coordinated parallel (see \autoref{fig:FoVs}). This parallel observation was designed to provide background comparison images by observing a field offset from the primary target by the intrinsic physical separation between NIRCam and MIRI on the JWST focal plane. We were able to position our pointing such that the eastern tile of our primary NIRCam mosaic was placed along the galaxy's main body, and the coordinated MIRI parallel was located north of the disk, offset from the Hubble X star-forming region. The MIRI parallel images were obtained at this location in the F1000W and F1500W filters using a \textsc{full} subarray, a \textsc{slowr1} readout pattern, with five groups per integration and one integration per exposure over the 12 total dithers, resulting in a total exposure time of 1433.4 seconds per filter (appendix \autoref{tab:obs_params_MIRI}). The total footprint encompasses 74\arcsec$\times$113\arcsec. This readout pattern was selected in contrast with the MIRI prime observations which used \textsc{fastr1} (see \autoref{sec:MIRI_strat}) due to data rate limits. 

One of our six mosaic tiles in the primary MIRI observation has a NIRCam coordinated parallel (second from the top; see  \autoref{fig:FoVs}). This tile was selected due to the advantageous position in the instrument focal plane of the Hubble IV star-forming region, which falls into NIRCam's Module B. Module A is situated further from the galaxy and is thus a useful pointing for obtaining information on a region of the galaxy that is not actively forming stars. The two regions image an area of 132\arcsec$\times$132\arcsec each. These parallels are imaged in the F140M, F150W, F277W, and F335M bands. A complete summary of the NIRCam parallel observing parameters including read mode, groups, integrations, and total exposure time is given in appendix ~\autoref{tab:obs_params_NIRCam}.

%%%%%%%%%%%%%%%%%%%%%%%%%%%%%%%%%%%%%%%%%%%%%%%%%%
%SECTION 3: DATA-PROCESSING APPROACH
%%%%%%%%%%%%%%%%%%%%%%%%%%%%%%%%%%%%%%%%%%%%%%%%%%

\section{Data Reduction and Photometry} % Section 3.
\label{sec:photometry}

\subsection{NIRCam Image Processing} % Section 3.1

The uncalibrated NIRCam images were processed through all three stages of the JWST pipeline using version 1.9.6 and Calibration Reference Data System (CRDS) version 11.16.20. For Stage-1 processing, we used CRDS context 1063 (\texttt{jwst\_1063.pmap}). We implemented the frame0 correction to recover stars that saturated in the first group ($\sim$21 s), but were unsaturated in the first 10.7 seconds comprising frame0 (\texttt{ramp\_fit.suppress\_one\_group=False}). Stage-2 processing was run with CRDS context 1075 (\texttt{jwst\_1075.pmap}). The output of Stage-2 (*\_cal.fits) were corrected for 1/f noise using the {\sc image1overf.py} tool from \citet{1fcor}. The resulting files were then aligned to {\em Gaia} DR3 using the {\em JWST/Hubble} Alignment Tool \citep[JHAT;][]{jhat}. Stage-3 mosaics were created with CRDS context 1077 (\texttt{jwst\_1077.pmap}). We skipped the tweakreg step in Stage-3 since the WCS alignment was provided already by JHAT. The difference in pmap for each stage of the pipeline occurred due to a rapid succession of reference file deliveries while we were processing the data. The only update relevant to this dataset between the 1063 and 1075 pmaps is an update to the snowball correction step. Snowballs are rare in this dataset, so we did not apply that correction.

\subsection{MIRI Image Processing} % Section 3.2

A description of the MIRI processing for this observational program is given in \citet{lenkic2023} but we give a short overview here and describe the treatment of the background in more detail. To create calibrated MIRI images we used JWST pipeline version 1.9.5 with CRDS version 11.16.21 and context \texttt{jwst\_1084.pmap}. Each of the raw MIRI files was processed through \texttt{Dectector1Pipeline} and the output through \texttt{Image2Pipeline} with default parameters. The resulting images were then aligned to \emph{Gaia}~DR3 using the \texttt{tweakreg} step in the pipeline. 

\begin{figure*} % Figure 2
    \centering
    \includegraphics[trim=0cm 0cm 0cm 0cm, clip=true, width=1.\textwidth]{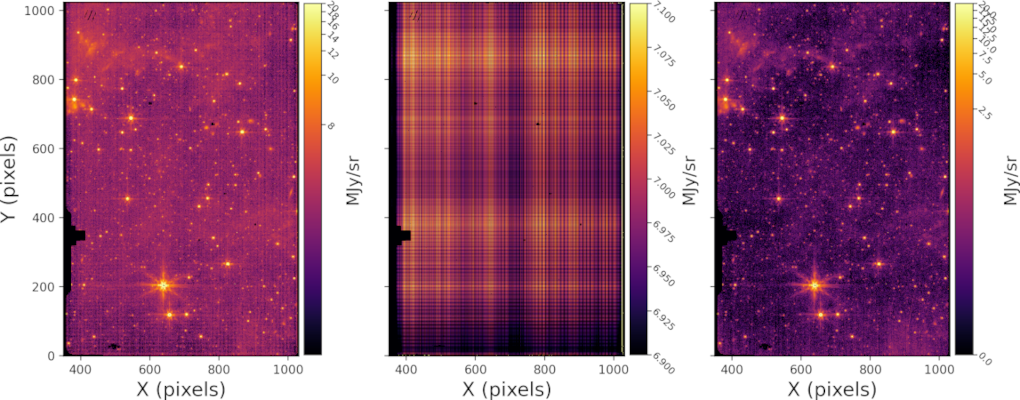}
    \caption{Example of background model applied to an F770W dither. The uncorrected image, the model instrumental background, and background subtracted images are shown in left, middle, and right, respectively. Only the main imager field is shown as we exclude the Lyot coronagraph from our mosaics. }
    \label{fig:f770w_bkg}
\end{figure*}

We generated instrumental backgrounds from Visits 001 and 005 of Observation 007 (i.e., the two tiles at the mosaic edges), as these are least affected by real diffuse emission. While this is not the ideal strategy, it helps to mitigate detector effects and reveal fainter sources in the galaxy. To generate an instrumental background image free of point sources and diffuse emission, we median combined all dithers of a given filter in both tiles. However, in the case of the F770W and F1000W filters, significant structure due to the extent of real diffuse emission in these tiles rendered these backgrounds unusable. For these filters we instead created a model instrumental background by taking comparatively unaffected detector rows and columns in the median combined background in order to create a representative median row and column. These were then mapped onto the detector plane to produce the model background which also accounts for row and column calibration artefacts \citep{Dicken2024}. An example of this is shown in ~\autoref{fig:f770w_bkg}. 

For the F1500W and F2100W filters, only minimal structure around the brightest sources remained in the median background. Even though the residual contamination is low, it can still manifest in background subtracted images as faint shadows dispersed on the image. To remove the residuals we manually masked their location and applied an approximate `filling' of the gaps using the \texttt{ndimage.distance\_transform\_edt} module in SciPy. Once backgrounds were created for all filters, we subtracted the instrumental background images from each dither in the mosaic tiles. Since Visit 001 of Observation 009 was performed 10 days after the other visits, we found that the background levels had increased slightly and corrected for this by determining a simple background offset from regions on the detector free of diffuse emission. Overall we found that our background subtraction methods performed well in reducing the background level across all mosaic tiles in all filters. We then constructed mosaics from the background-subtracted images using \texttt{Image3Pipeline}.

\subsection{Source Detection} % Section 3.3
\label{sec:sourcedetection}
Point-source photometry was extracted using the {\sc starbugii} \citep{Starbug2} photometric tool and pipeline which is optimised for JWST observations of crowded stellar populations within complex environments \citep[e.g.][]{Jones2023_NGC346}. This pipeline performs point-source extraction and band merging across multiple observations and wavelengths utilising core functions from the python {\sc photutils} \citep{larry_bradley_2022_6825092} package. A complete set of relevant {\sc starbugii} parameters and their values adopted for our photometric extractions are listed in appendix ~\autoref{tab:sb_params}. 

The individual exposures from the {\em Gaia}-aligned Stage-2 
images are used for source detection with \texttt{starbug2 $--$detect}. Sources with a \texttt{$5\sigma$} detection above the sky, which is initially estimated locally within an annulus, are first located by centroid fitting. The geometry of each point source is calculated and {\em sharpness} and {\em roundness} values are assigned. {\em Sharpness} is the ratio of the peak pixel value of the source to its median within an aperture, we remove cosmic rays from the catalogue by setting an upper limit on {\em sharpness} and faint peaks in the dust structure by setting a lower limit. The source symmetry is measured by comparing normal functions fit to the source along the vertical and horizontal axis, this is defined as{\em roundness}; we are able to remove most resolved background galaxies and further spurious detections within the dust structure by limiting the allowed level of asymmetry. 

The NIRCam {\sc fullbox 4tight} dither mode results in four sets of three overlapping exposures for a given pointing. Only sources detected in at least two frames of each set are retained in the resulting catalogue; any sporadic source not meeting this threshold is discarded as a likely cosmic ray or other detector artefact while retaining true sources at the detector edges. For the MIRI images, the {\sc full cycling, small} mode results in four dithers, so we stipulate that a source must be detected in three or more exposures. The flux distribution between the exposures is examined and sources with asymmetric distributions (the mean and median differ by more than 5\%) are assigned the flag \texttt{SRC\_VAR}. Matching between frames in the image stack is done with a nearest neighbour calculation with a threshold separation of 0.1 arcseconds. 

We conduct aperture photometry on these sources in each individual frame.
An aperture radius scaled to 60\% of the encircled energy of the PSF and a sky annulus with an inner and outer radius of 3.0 and 4.5 pixels, respectively, is used for all NIRCam bands, and the aperture correction is interpolated between values given in CRDS \texttt{jwst\_nircam\_apcorr\_0004.fits}. The MIRI images are treated the same, with 60\% encircled energy aperture sizes but we increase the sky annuli to 4.0-5.5 pixels for F770W and F1000W, 5.0-6.5 pixels for F1500W and 6.0-7.5 pixels for F2100W to account for the larger PSFs at longer wavelengths. Aperture corrections are calculated from \texttt{jwst\_miri\_apcorr\_0005.fits}.
The data quality array within the aperture of each source is inspected and we flag sources with saturated or \texttt{DO\_NOT\_USE} pixels with the {\sc starbugii} flag \texttt{SRC\_BAD} and pixels that contained a jump during detection with \texttt{SRC\_JMP}.

\autoref{tab:sensitivity} contains a list of source counts for every filter and an estimate of their sensitivity values.

\subsection{PSF Photometry} % Section 3.4

We conduct PSF photometry for the sources in the NIRCam images.

The nebulous background emission underlying our images of NGC~6822 is modelled using \texttt{starbug2 $--$background}. This routine places masking apertures of varying sizes, calculated on the flux of each source, and fills them with the median pixel value within a local annulus. For every pixel in the image, the background is measured by averaging all the local pixels within a set box size. This process creates an effective representation of the nebulous emission; 
by subtracting this diffuse emission background from a single exposure, we create a clean image of our field with nebulous emission removed onto which PSFs can be accurately fit.

For each detector subarray within NIRCam, we generated a five arcsecond PSF with {\sc webbpsf}~\citep{webbpsf2014} version 1.1.1. The single exposures have their nebulous background estimated and subtracted, and we run PSF photometry using \texttt{starbug2 $--$psf} on the residual image at the source positions from the combined and cleaned source list generated in our aperture photometry step. The centroid position is left as a free parameter, allowing both flux and position to be fit during the routine. If the newly fitted centroid position differs from the initial guess by greater than 0.1 arcseconds, we refit the flux but hold the position fixed at the initial guess. This results in a poorer fit and therefore we flag the source with \texttt{SRC\_FIX}.

\subsection{Photometric Corrections} % Section 3.5
\label{sec:corrections}

We calculate and apply instrumental zero point magnitudes to calibrate the PSF photometry because they are not normalised to physical units. For each filter, a cleaned aperture photometry catalogue which retains only the most reliable point sources is used as the base to determine the zero point.
To produce the clean source list from the main catalogue, both the faintest and brightest sources are removed to limit sources with low S/N, partially saturated objects, and any potential remaining detector artefacts. Sources must have a photometric error of less than 0.1 and must not have any poor-quality data flags. These cuts eliminate over 80\% of the sources in the main catalogue. Finally, these sources are matched to the equivalent source in the PSF catalogue. We use \texttt{starbug2 $--$calc-instr-zp} to calculate the median difference in the source magnitudes measured by aperture photometry and PSF fitting to obtain instrumental zero point magnitudes which are subsequently used to calibrate the PSF photometry from {\sc starbugii} to the AB magnitude system.
For easier comparison with past works, we convert these magnitudes to the Vega system using the reference files \texttt{jwst\_nircam\_abvegaoffset\_0001.asdf} and \texttt{jwst\_miri\_abvegaoffset\_0001.asdf}.

Finally, as NGC~6822 is located at a low galactic latitude it is necessary to correct the NIRCam photometry for foreground reddening. We adopt the value E(B$-$V)\,$=0.35$~\citep{Tantalo_2022} to correct for the moderate Galactic foreground extinction and apply the extinction curve of \cite{Cardelli1989} assuming $R_V = 3.1$. 
No extinction corrections were applied to the MIRI data, as foreground reddening in mid-IR wavelengths is negligible. Nor was differential extinction internal to NGC~6822 accounted for as this is assumed to be insignificant compared to photometric uncertainties due to crowding. 

\subsection{Catalogue} % Section 3.6

Individual filter band catalogues are merged together to form a combined NIRCam and MIRI point-source catalogue for NGC~6822 with \texttt{starbug2-match $--$band}. Prior to any matching, the photometric uncertainty of each source is assessed, and anything with a magnitude error greater than 10\% is removed from the individual catalogues. This reduces the likelihood of mismatching high-quality sources with any remaining spurious sources, producing a reliable band-matched catalogue at the possible cost of completeness.  

Initially, we treat the NIRCam and MIRI catalogues separately, applying the same methodology to the band matching within each instrument.  
The matching routine starts with the shortest-wavelength catalogue, as the smaller PSF full width at half maximum (FWHM) leads to the highest astrometric certainty. This is matched with a nearest neighbours method to the next-shortest wavelength catalogue, with any unmatched sources above the separation threshold appended to the end. The position used for each source is taken from the shortest-wavelength catalogue in which it initially appeared, allowing for faint red objects such as highly dust-enshrouded AGB stars that are not visible in the near-IR to be retained.
We used a separation threshold that increases as the PSF size increases, as using a single threshold larger than the astrometric uncertainty of the longest MIRI wavelengths would cause mismatching in the shortest NIRCam filters. We match short wavelength NIRCam filters using a threshold of 0.06\arcsec{} and long wavelength NIRCam filters with 0.1\arcsec{}. 
Within MIRI we adopt 0.15\arcsec{} for F770W and F1000W, 0.2\arcsec{} for F1500W, and 0.25\arcsec{} for F2100W.

Finally, we merge the NIRCam and MIRI catalogues with a separation threshold of 0.3\arcsec. We expect many of the reddest and most dust-enshrouded sources detected in the long-wavelength MIRI data to be missing from the shortest-wavelength NIRCam images. However, the depth and sensitivity of F115W and F200W result in a catalogue populated by many faint blue sources. Consequently, a simple positional matching approach between a long-wavelength MIRI band and a short-wavelength NIRCam band would likely result in a significant number of mismatches because of the many sources present at short wavelengths which lie along similar lines of sight. To combat this issue we require any sources matched between NIRCam and MIRI to have an F444W detection. In other words, a source in our MIRI catalogue will be compared with all sources in the NIRCam catalogue within the matching radius and paired with the nearest source for which a detection in the F444W band exists. If no NIRCam sources within the matching radius are detected in the F444W band, the MIRI source will be assumed to have no NIRCam counterpart and will subsequently be appended to the catalogue as a new object. Thus both blue and red objects in the catalogue will be retained and the chance of a mismatch between co-located sources reduced.

\subsection{Foreground Contamination} % Section 3.7.1
\label{sec:FGDcontamination}

To remove foreground Milky Way stars from our JWST NGC~6822 source list we examine {\em Gaia} Data Release 3~\citep[DR3][]{GAIAeDR3} for foreground contamination. First, the {\em Gaia} catalogue is cleaned of sources with poor astrometry and possible non-single objects within our JWST FoV, as outlined by \cite{Fabricius2021b}. This involves keeping sources with:
RUWE$\le$1.4,\ 
astrometric\_excess\_noise\_sig$\le$2.0,\
visibility\_periods\_used$\ge$9 and\
ipd\_gof\_harmonic\_amplitude$\le$0.1. 
The cleaned {\em Gaia} catalogue is matched to our NGC~6822 photometry, resulting in 656 positive matches within the main field. Sources that exhibit a significant ($5\sigma$) parallax or proper motion are considered to be foreground sources and are thus removed from our catalogue. Due to the relative proximity of NGC~6822, we check that the proper motion of the sources do not deviate significantly from the global proper motion of the galaxy, as outlined by \cite{Dimitrova2021}.

In total, we find that 179 sources exhibit proper motion above the threshold with all but one also deviating significantly from that of the galaxy. We also find that every source with significant parallax also exhibits proper motion. This allows for 178 sources to be removed from the catalogue as foreground contaminates.
Despite NGC~6822's proximity to the galactic plane, we see a low number of foreground stars in our data set. Our detection routine is unable to recover many of the bright foreground stars due to the saturation limits of our data. Furthermore, \emph{Gaia} only detects the brightest IR sources in NGC~6822; this paired with the small FoV of our observations means that the number of foreground contaminating stars in our catalogue is low, as expected.

%%%%%%%%%%%%%%%%%%%%%%%%%%%%%%%%%%%%%%%%%%%%%%%%%% 
%RESULTS
%%%%%%%%%%%%%%%%%%%%%%%%%%%%%%%%%%%%%%%%%%%%%%%%%%

\section{Results and Discussion} % Section 4
\label{sec:prelimresults}

\subsection{Mosaic Images} % Section 4.1.
\label{sec:images}

\begin{figure*} % Figure 3
    \centering
    \includegraphics[width=\textwidth]{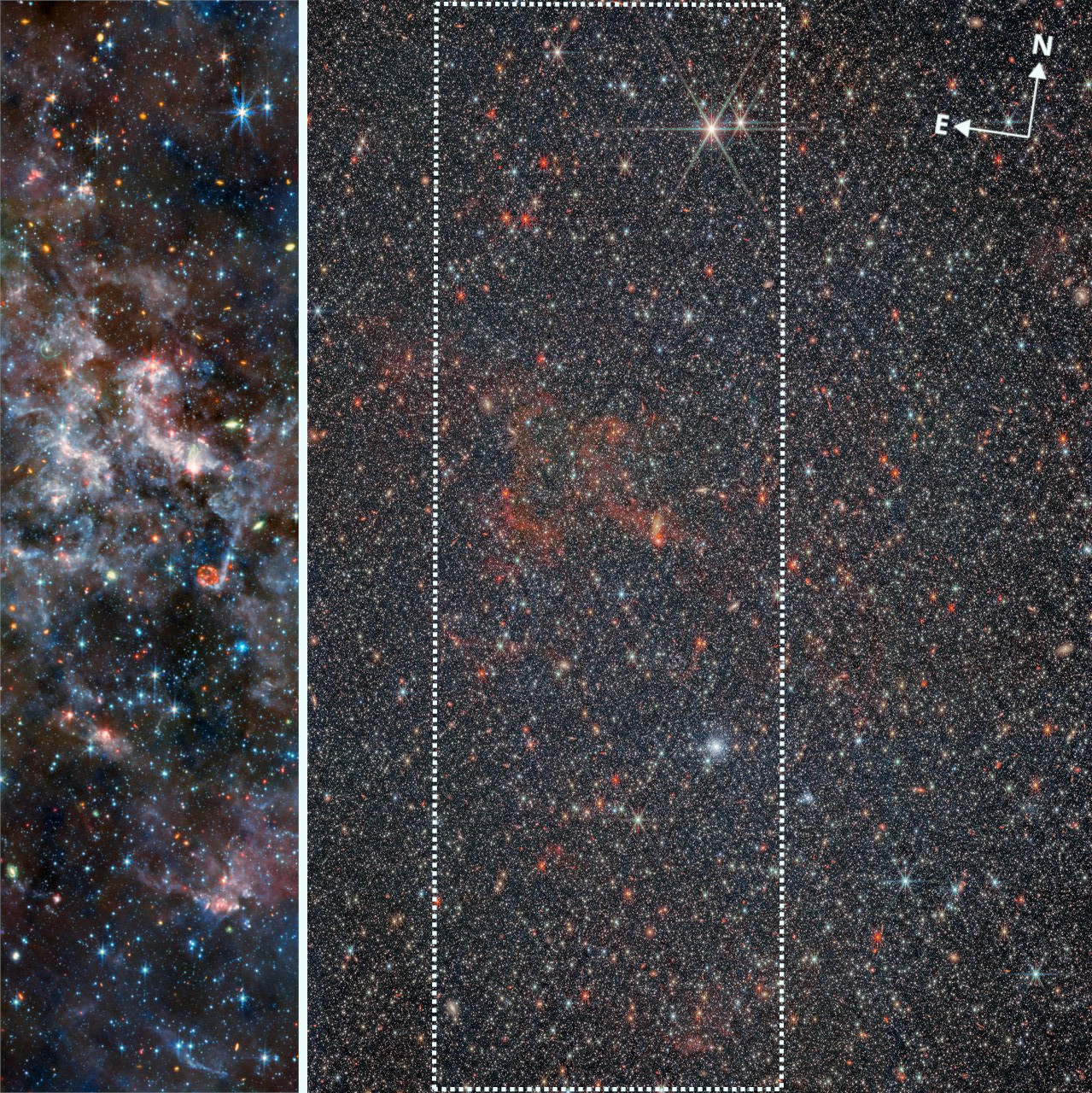}
    \caption{JWST full-colour mosaics displaying the spatial coverage from both instruments. Left shows MIRI coverage with F770W (blue), F1000W (green), F1500W (yellow) and F2100W (red). Right shows the NIRCam coverage with F115W (blue), F200W (cyan), F356W (yellow) and F444W (red) with the extent of the MIRI footprint marked with the white dotted line. East is to the left and north points upward. Images credited to ESA/Webb, NASA \& CSA, M.\ Meixner.}
    \label{fig:colourImage}
\end{figure*}

The NIRCam prime mosaics shown in ~\autoref{fig:colourImage} cover a FoV of $\sim$29.0 arcmin$^{2}$, whilst the MIRI mosaic covers a $\sim$14.5 arcmin$^{2}$ FoV, with the NIRCam FoV overlapping nearly the entirety of that of MIRI.
These mosaics image NGC 6822, located at a distance of 490~kpc, with resolutions from 0.040\arcsec{} (0.095 pc) to 0.145\arcsec{} (0.344 pc) in NIRCam, and from 0.268\arcsec{} (0.638 pc) to 0.673\arcsec{} (1.601 pc) in MIRI. This provides an improvement over existing \emph{Spitzer} imaging by up to a factor of 10.

The NIRCam field shows a smooth distribution of stars increasing in density towards the centre of the FoV. Scanning by eye reveals an assortment of clear extreme candidates with bright blue or red colours on a backdrop of many faint stars. The background galaxies are evenly distributed across the image, with some showing detailed spiral structures and others only visible in the longest wavelengths. A dense globular cluster (Hubble VII) is visible in the centre-right of the image. 
The extent of the MIRI footprint is overlaid as a white dotted line. The NIRCam data are almost devoid of all dust structure whereas the MIRI data of the same region are completely dominated by complex diffuse emissions. The centre of our FoV covers the young massive star-forming region Spitzer~I, for which a detailed study of the young stellar objects has been conducted by \citet{lenkic2023}.

\subsection{Stellar Population Classification} % Section 4.2.
\label{sec:soclassapp}

This section presents JWST CMDs covering the galaxy within the full extent of our survey. We select colour combinations that highlight significant population separation and those which represent similar wavelength coverage as \emph{Spitzer} filters used in the SAGE surveys~\citep{bib:Meixner2006,bib:Gordon2011}. We adapt point-source classification methodologies and boundaries from prior work ~\citep{bib:Blum2006,bib:Jones2017,bib:Hirschauer2020} and investigate the location various stellar populations occupy in different JWST colour combinations, which we anticipate will be useful in future stellar surveys.

\begin{figure*} % Figure 4
    \centering
    \includegraphics[width=0.45\textwidth]{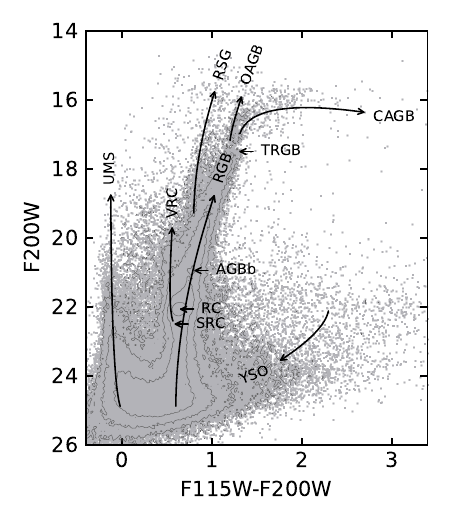}
    \includegraphics[width=0.45\textwidth]{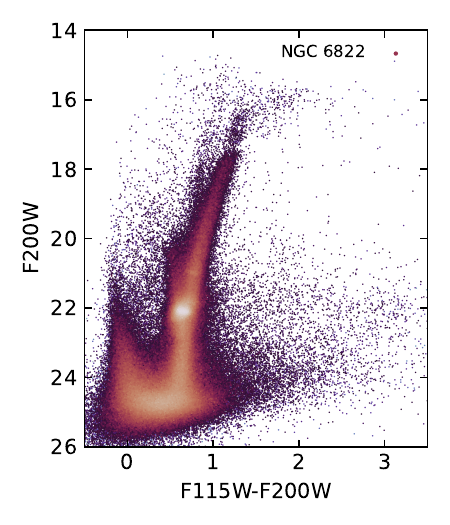}
    \includegraphics[width=0.45\textwidth]{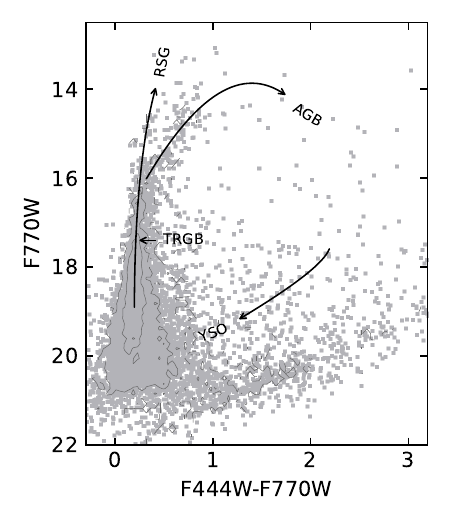}
    \includegraphics[width=0.45\textwidth]{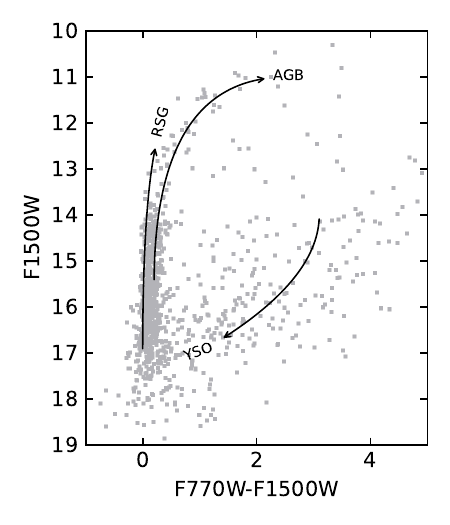}
    \caption{Colour-magnitude diagrams over three filter combinations with broad stellar classification tracks overlaid and labeled accordingly, point density displayed as contours where appropriate: $\mathrm{F115W}-\mathrm{F200W}$ vs.\ $\mathrm{F200W}$ (top), $\mathrm{F444W}-\mathrm{F770W}$ vs.\ $\mathrm{F770W}$ (lower left) $\mathrm{F770W}-\mathrm{F1500W}$ vs.\ $\mathrm{F1500W}$ (lower right). Upper right plot shows a second $\mathrm{F115W}-\mathrm{F200W}$ vs.\ $\mathrm{F200W}$ CMD in Hess format.}
    \label{fig:roadmap}
\end{figure*}

\autoref{fig:roadmap} shows one NIRCam-only CMD (F115W$-$F200W vs.\ F200W) in contour and Hess diagram format, one MIRI-only CMD (F770W$-$F1500W vs.\ F770W), and one combined CMD (F444W$-$F770W vs.\ F770W) with the most prominent populations overlaid. The first CMD, with a filter combination most similar to the ($J-K_s$ vs.\ $K_s$) CMDs of previous NGC~6822 studies (e.g., \citealp{bib:Cioni2005,bib:Sibbons2012, bib:Whitelock2013}), was chosen as it utilizes the most sensitive photometry and therefore reveals the faintest populations. The diagram is shaped as a collection of sources that splits vertically into two main fingers from the base, with the broadness of the base likely being a product of greater scatter as the sensitivity limit of the photometry is approached. The blue finger follows the Upper Main Sequence (UMS), with more massive stars appearing higher up. The right-hand fork is primarily the Red Giant Branch (RGB), with the Red Clump (RC) appearing as a dense bulge at F200W$=$22.1 and the AGB bump (AGBb) embedded just above at F200W$=$21.0. The vertical track seen above the RC is formed by intermediate-mass helium-burning stars (VRC) evolving toward the AGB. The red supergiant (RSG) track splits from the RGB below the Tip of the RGB (TRGB) and runs diagonally upward, consisting of some of the brightest sources in the galaxy. Above the TRGB the thermally pulsing AGB (TP-AGB) stars separate into oxygen-rich AGB (OAGB) and carbon-rich AGB (CAGB). A fraction of the TP-AGB will likely be mid-thermal pulse and sit below the TRGB. Formal separation of these fainter AGB stars from RGB stars is difficult photometrically due to this overlap, however, they make up only a small percentage (${<}10\%$) of the overall TP-AGB population \citep{bib:Boyer2015a}.

\subsubsection{Upper Main Sequence}\label{sec:class:UMS} % Section 4.2.1

The upper main sequence population is visible in the top left panel of \autoref{fig:roadmap} as the prominent branch located at $\mathrm{F115W}-\mathrm{F200W}<0.4$ and approaching  $\mathrm{F115W}-\mathrm{F200W}=0.0$.  
At $\mathrm{F200W}=25.0$ this branch connects with the base of the RGB, though it is unclear if this constitutes the Main Sequence Turn-Off (MSTO) for an aged population as the confusion introduced into the photometry as it approaches the completeness limit merges the populations together.
An optical survey of NGC 6822 by \citet{Zhang2021} isolated the UMS with ages $<$100~Myr and confirmed that the population perfectly traced the \HI{} gas distribution that lies orthogonal to the bulk of the stellar component of the galaxy. Because they trace recent star formation, these stars offer information about the underlying gas and how it evolves over time.

UMS stars are generally very bright and have dispelled any nearby dust, they possess spectral energy distributions (SEDs) peaking in blue wavelengths. As such, they are detectable in the short NIRCam filters but become difficult to detect at longer wavelengths. 

\subsubsection{Red Giant Branch} % Section 4.2.2

The RGB begins at $\mathrm{F200W}=25$ and $\mathrm{F115W}-\mathrm{F200W}=0.6$ in the top left panel of \autoref{fig:roadmap} and extends vertically in a steep diagonal towards the red. This sequence is formed by stars that have just left the MS and have begun H-shell burning. It is one of the most prominent features in near-IR CMDs. In a single-aged population, the MSTO will appear as a tight track connecting the MS to the base of the RGB. The multi-aged population of NGC~6822, a product of continuous star formation, elongates the RGB and is consistent with an old population spanning $2-10$~Gyr.
Previously \citet{Zhang2021} isolated the brighter portion of this branch to show that the old stars contribute to the galaxy's eccentric elliptical stellar component that twists radially and is perpendicular to the younger stellar component. 

The TRGB is well defined in the shorter wavelengths in \autoref{fig:roadmap} but unresolved in the longer MIRI filters. Our method of roughly calculating its position is discussed in Section~\ref{sec:lfs}.

The positioning of the RGB on a given CMD is fairly independent of star formation history (SFH) but two effects cause it to spread in colour space: stellar age and metallicity. Younger RGB stars appear slightly to the blue, while older RGB stars appear more to the red when considering a fixed value of metallicity. Whereas when given a fixed population age with varying metallicity, more metal-poor stars appear bluer than the more metal-rich stars. This effect is known as the age-metallicity degeneracy~\citep{Worthey1999,Carrera2008} as for a given individual star, it is difficult to determine with certainty if its position on the RGB is caused by its age, metallicity or both. It results in a spreading of the sequence, the width of the RGB in our CMDs exceeds the expected characteristic scatter caused by uncertainty in the photometry and therefore is likely caused by a broad range of ages and metallicities across the galaxy. 
Indeed, detailed surveys of the metallicity of NGC 6822 have been conducted by \citet{Patrick2015}, who detect a metallicity gradient across the innermost regions of the galaxy, which coincides with our JWST FOV.
Furthermore, the spread in RGB colour in the JWST CMDs is in agreement with \cite{Tantalo_2022}, who find that the RGB in NGC 6822 spans 1.2 mag in $g-i$ colour, indicative of NGC 6822's complex star-formation history and chemical composition.

\subsubsection{Red Clump}\label{sec:rc} % Section 4.2.3

Low mass stars evolving up the RGB deposit helium onto their degenerate core, once a sufficient core mass \citep[$0.48M_\odot$;][]{Bildsten2012} is reached the core temperature breaks the electron degeneracy suddenly, in a helium flash \citep[][and references therein]{AGB_Book}. 
The uniform initial characteristics of the now helium-burning cores leads to the stars collecting in a tight sequence on the CMD, the red clump (RC), with the differing properties largely affecting their convective outer layers. Lower mass or more metal-poor stars will have less massive convective layers above the core which will cause less ``dampening'' of the radiation and therefore appear in the hotter horizontal branch (HB), not immediately visible on our CMDs~\citep[see][for a detailed review on RC stars.]{Girardi2016}.

Due to its prominence and well-defined position in CMDs the RC can be used as a standard candle due to its small dependence on age and stellar composition~\citep[e.g.,][]{Bovy2014, RuizDern2017, Ting2018, Hawkins2018} although there is a small metallicity dependence~\citep{Alves2000,Shan2018}.

The over-density from the RC structure is pronounced in our CMDs, sitting just left of the RGB at $\mathrm{F200W}\sim22.1$. In our shortest-wavelength filters we resolve sources three magnitudes below the RC, but a decrease in the depth of photometry at redder wavelengths prevents its detection at $\lambda >$ 4~$\mu$m, as seen in~\autoref{fig:roadmap}.
The RC has been used to measure SFH in NGC~6822 \citep[e.g.,][]{Wyder2001} by comparing age-sensitive RGB with the age-insensitive RC. The increased sensitivity of our catalogue will allow for more finely sampled calculations of the SFH within the inner regions of the galaxy. 
Additionally, differential reddening in NGC 6822 introduces a spread in the photometry of the theoretically tight RC. This effect can be used to build an extinction map of the galaxy~\citep{Wyder2001}. Such an exercise is beyond the scope of this study, however, it will be addressed in a forthcoming paper.

\subsubsection{Vertical Red Clump} % Section 4.2.4
\label{sec:vrc}

Intermediate mass stars do not experience a helium flash and instead begin quiescently burning helium in their cores~\citep{AGB_Book,SNHandbook2017}. The ignition of helium moves them off the RGB and onto a vertical sequence shown on the CMDs in the upper panels of \autoref{fig:roadmap} at a colour of F115W-F200W=0.6 and 2.5 magnitudes from F200W=22.5. The feature is often associated with the RC on CMDs and therefore gets the name, Vertical Red Clump (VRC)~\citep{Girardi2016}.

VRC star's luminosity monotonically increases with stellar mass~\citep{Girardi1999}. The feature extends a few tenths of a magnitude below the RC, with the limiting depth marking the critical lowest mass above which helium can ignite quiescently in the core \citep[$\sim2M_\odot$;][]{Chiosi1992}. This is visible as a small flume underneath the RC in upper right panel of \autoref{fig:roadmap} and is known as the Secondary Red Clump \citep[SRC;][]{Girardi1998}.

\subsubsection{AGB Bump}\label{sec:agbbump} % Section 4.2.5

The AGBb is a population of rarely-observed stars that emerges as evolving early AGB stars stall momentarily as their H-burning shells are extinguished by the expansion of their convective layers, with the ignition of the He-burning shell \citep{Caputo1978, Dreau2022}. Observing this feature is difficult, as it requires a large number of sources and may be quickly obscured by even small photometric errors. Yet, it is clearly visible in the upper plots of \autoref{fig:roadmap} embedded in the RGB at $\mathrm{F200W}=21$. 
The age and metallicity of the stars influence its positioning on the CMD and as such, 
the AGBb can be used to gain some insight into the SFH of the galaxy~\citep{Ferraro1992}. 
Although the AGBb is visible in optical data in NGC~6822 ~\citep{Tantalo_2022}, this often neglected population is a useful mark to constrain evolutionary tracks.

\subsubsection{Red Supergiants} % Section 4.2.6

At $\mathrm{F200W}=19.5$, the RSG branch forks blueward from the RGB and extends steeply to the saturation limit of our data. The sequence is clearly visible in the upper-right plot of \autoref{fig:roadmap}, although the source density drops at the brightest end. This corresponds to the upper part of the red helium burning sequence.
The existence of a RSG population in the galaxy has been shown by \citet{bib:Whitelock2013}, \citet{bib:Hirschauer2020}, \citet{Dimitrova2021} and \citet{Tantalo_2022}, where they follow the structure of the central bar. This result was expected as RSG stars are young (8--20 Myr) and massive ($\sim$8--40 M$_\odot$), so have not had sufficient time to wander far from the star-forming sites in the centre of the galaxy \citep[e.g.,][]{Humphreys1979, Massey2003, Davies2018}. 
The colour-cut to isolate this branch is denoted by the green region drawn in the left panel of ~\autoref{fig:cmd:evolv}. The region is drawn by eye based on point densities in the CMD and was informed by previous NGC 6822 stellar classification studies \citep{bib:Sibbons2012,bib:Hirschauer2020}.
The isolated RSG stars form a distinct blue ``horn'' in $\mathrm{F444W}-\mathrm{F770W}$ vs.\ $\mathrm{F770W}$ as seen in the lower-left panel of ~\autoref{fig:roadmap} where the labelled arrow marks the span of the branch. In the longer MIRI wavelengths, the generally dust-free stellar sources compress at a colour of zero, whereas RSG and O-rich AGB stars that are surrounded by dust envelopes also have similar, but redder, mid-IR colours due to their dust grains following the same condensation sequence \citep{Ferrarotti2006}. Thus separating RSG and AGBs (with and without a circumstellar excess) becomes more difficult in the lower right panel of ~\autoref{fig:roadmap}.

\subsubsection{Asymptotic Giant Branch} % Section 4.2.7

\begin{figure*} % Figure 6
    \centering
    \includegraphics[width=0.49\textwidth]{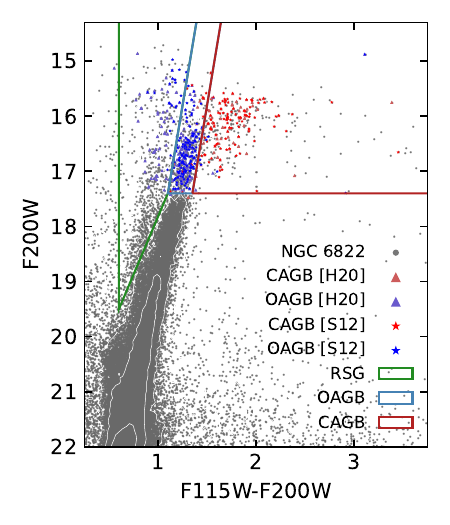}
    \includegraphics[width=0.49\textwidth]{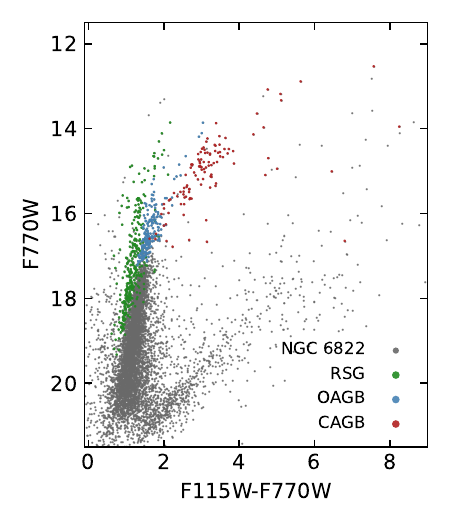}
    \caption{CMDs showing the evolved population separation from near-IR to mid-IR. The candidate RSGs in green, OAGBs in blue, and CAGBs in red in the right panel are defined by the solid boundaries shown in the $\mathrm{F115W}-\mathrm{F200W}$ vs.\ $\mathrm{F200W}$ CMD (left panel).  The white contours represent the source density on the RGB.  The left plot also shows the OAGB (blue) and CAGBs (red) candidates classified by \citet{bib:Sibbons2012} (stars) and \citet{bib:Hirschauer2020} (triangles). 
    }
    \label{fig:cmd:evolv}
\end{figure*}

Here we present initial cuts to determine the general positions of the RSG, OAGB, and CAGB stars. To inspect the locations of these evolved populations in colour space, we show in the first panel of \autoref{fig:cmd:evolv} a zoom-in of the F115W$-$F200W vs.\ F200W CMD, where they appear to follow diagonal sequences upwards and to the red following the stars core H and He exhaustion. Thermally pulsing AGB stars occupy the reddest and brightest branches. Above the TRGB at $\mathrm{F200W}=17.5$, a small gap separates the TP-AGB stars from the RGB. For populations fainter than the TRGB there can be substantial overlap in colour between early AGB stars, RSGs and the more numerous RGB stars \citep[as seen by][in NGC 6822]{Tantalo_2022} limiting the classification accuracy. 

AGB stars can be separated by the distinct chemical makeup of their photospheres. Here, free carbon and oxygen are bound in very stable CO molecules. The overabundance of one of the two components is then left unbound in the photosphere and will form the basic ingredient for the atmospheric chemistry and dust grain formation once it is lifted from the stellar surface into the circumstellar envelope of the star \citep{Bowen1988}. AGBs with a $C/O$ ratio $> 1$ are defined to be C-rich and those with $C/O<1$ are O-rich \citep{Iben1983}. The molecular and dust species formed are dictated by this chemical difference, with CAGBs forming carbonaceous molecules and dust grains whilst OAGBs form oxides and silicates, enabling these evolved populations to be photometrically separated~\citep[e.g.,][and references therin]{AGB_Book, bib:Hofner2018}. 

For the bright TP-AGB stars, we separate the CAGB and OAGB sequences with a diagonal line above the TRGB at $\mathrm{F115W}-\mathrm{F200W}\ge1.4$ with a gradient of -12.8. Historically the AGBs in NGC~6822 have been split with a vertical line in the near-IR \citep[see e.g.,][]{bib:Cioni2005,bib:Sibbons2012}, 
or by employing a wide range of colour diagnostics to identify members of this bright intermediate-age stellar population \citep[e.g.,][]{Tantalo_2022}.
Combining mid-IR {\em Spitzer} and near-IR United Kingdom Infrared Telescope (UKIRT) data for broader baseline photometry, \citet{bib:Hirschauer2020} developed a novel statistical approach to separate them with more complex colour cut boundaries. We cross-match our catalogue with the catalogues of AGB candidates from \citet{bib:Sibbons2012} and \citet{bib:Hirschauer2020} to identify where these populations lie on our JWST CMDs. We show this comparison in the first panel of \autoref{fig:cmd:evolv}, where blue data points indicate OAGBs and red data points indicate CAGBs. 
We find that the classification of sources as CAGB stars by both of these previous studies are broadly consistent with the colour space that they occupy in our JWST CMDs (red region), with a total of 165 sources in agreement and 14 OAGBs being classified in our cuts as CAGB. 
Similarly we find a total of 241 OAGB classifications in agreement with 21 CAGBs being classified as OAGBs here.

Separating the OAGBs and RSGs is more involved due to the lower point density in our CMD at the brightest magnitudes. However, RSGs typically have warmer effective temperatures than AGB stars and do not experience a third dredge-up, so ordinarily have slightly bluer colours and are less numerous than the AGB stars. 
To select RSG stars, we adopt a diagonal line above the TRGB at $\mathrm{F115W}-\mathrm{F200W}=1.1$; whereas below the TRGB our colour selection is informed by the prominent density changes at the blue edge of the RGB branch, until a depth of 19.5 mag when the brightest VRC stars begin to overlap with RSG stars; this region (denoted in green) is where we classify objects as RSGs. In the Magellanic Clouds, the $J-K_s$ vs.\ $K_s$ CMDs of \citet{bib:Blum2006} and \citet{bib:Boyer2011} possess a similar RSG feature. 

The majority of the OAGB candidates from \citet{bib:Sibbons2012} and \citet{bib:Hirschauer2020} are classified as OAGB stars in our JWST data. However, 
70 of these sources appear to fall within the space we define as corresponding to RSGs. 
This is to be expected as, separating these classifications becomes difficult towards the mid-infrared~\citep{bib:Boyer2011, Jones2017_spec}. Here there is some overlap between the two branches as stellar temperature no longer affects the CMD; spectroscopy is required to definitively distinguish between them.
Finally, there are a few sources that were previously classified as OAGB or CAGB that fall along the RGB; these are likely TP-AGBs that are mid-pulse or stars that have not yet undergone a third dredge-up \citep{Iben1983}.
A caveat of this comparison is that the improved resolution of JWST resolves previous bright \emph{Spitzer} sources into multiple fainter stars. These would appear on the CMD in the first panel of \autoref{fig:cmd:evolv} to the right of the RGB and below the space occupied by CAGBs. These sources \citep[roughly half from the catalogues of][]{bib:Sibbons2012,bib:Hirschauer2020} have however been omitted from \autoref{fig:cmd:evolv} for clarity. 
While the colour cuts we have applied here effectively distinguish the brighter, least contaminated regions of the evolved stellar populations' colour space, we caution that circumstellar extinction imposes limitations on identifying the dustiest sources in the near-IR. Although it is assumed that everything to the red of the diagonal cut we have defined at F115W$-$F200W$=1.35$ (red line) is a CAGB~\citep{bib:Boyer2015b}, in cases where OAGBs and RSGs produce large quantities of silicate dust, they will also occupy this colour space~\citep{Aringer2016, Jones2017_spec}. Separating the dustiest OAGB and CAGB stars requires spectroscopic observations.
We also expect that for the most extreme dust-enshrouded AGB stars, extinction will cause the star to have a very red colour and be fainter than the TRGB used to identify the evolved stars in \autoref{fig:cmd:evolv}. As such, these heavily extinguished sources will be excluded from our AGB classification. However, these objects are extremely rare and short-lived \citep{vanLoon2010}.

Using the classification schemes defined above, we overlay the sources onto F115W$-$F770W vs.\ F770W in the second panel of ~\autoref{fig:cmd:evolv}. The RSGs, OAGBs, and CAGBs are still visible as three distinct populations, although increasing absorption effects caused by dust in the longer wavelengths begin to increase the scatter of the populations. This colour combination separates the CAGBs from the OAGBs, with the bulk of the two populations being separated with nearly one magnitude of colour space between them. This combination holds close comparison with the $J-[8.0]$ CMD of the LMC from \cite{bib:Blum2006}, where the RSGs and CAGBs fork above the bulk of OAGBs; with NIRCam- and MIRI-equivalent filters we see the same morphology. 

\begin{figure}
    \centering
    \includegraphics[width=0.45\textwidth]{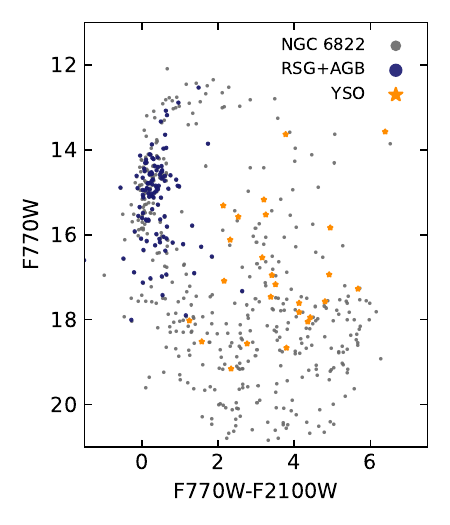}
    \caption{Colour-magnitude diagram with MIRI $\mathrm{F770W}--\mathrm{2100W}\ vs.\ \mathrm{F770W}$. NIRCam identified RSG, OAGB and CAGB candidates are overplotted in blue and YSO sources from \citet{lenkic2023} are overplotted in orange.}
    \label{fig:evolvedmiri}
\end{figure}

Long-wavelength MIRI photometry is key to chemically separating the most dust-enshrouded sources in our catalogue. \autoref{fig:evolvedmiri} shows the distribution of sources in $\mathrm{F770W}-\mathrm{F2100W}$ vs.\ F2100W, where the source density is very low due to the significant dust presence and rarity of sources which are bright at these wavelengths.
The CMD splits into two distinct groups stretching across six magnitudes in colour space. This combination has shown to be analogous to \emph{Spitzer} $[8.0]-[24]$ vs.\ $[24]$ in work by \citet{bib:Jones2017}, which allows us to draw a comparison to SAGE work on the LMC where \citet{Srinivasan2009} demonstrated a triple forking sequence. In their work, CAGB (and extreme-AGBs) lie on the most-luminous and blue finger of the fork, whereas the OAGBs are fainter and redder but bifurcate into two fingers above $[8.0]-[24]>1$. Later, \citet{Sargent2011}, using the Grid of Red supergiant and Asymptotic giant branch star ModelS (GRAMS) models, showed that the central fork is occupied by the OAGBs exhibiting the highest mass loss as well as some RSG stars. 
Our CMD suffers from a low number of sources, drawing a direct comparison is difficult but the most extreme AGB stars turn over at F770W$\sim$13 and the RSG and less extreme AGBs collect on F770W-F2100W=0.

\subsubsection{Young Stellar Objects} % Section 4.2.8

In all filter combinations shown in \autoref{fig:roadmap}, the YSOs inhabit the red area to the right of the RGB below the CAGB sequence. These red dust enshrouded objects have characterisable SEDs which typically rise towards longer wavelengths. \citet{lenkic2023} identifies 140 YSOs in the Spitzer~I region in NGC 6822, using JWST data. We overlay the high confidence YSOs from that study onto \autoref{fig:evolvedmiri}, demonstrating the effective separation of these young pre-main sequence sources from AGB and RSG sources using MIRI filters.  The JWST colour cuts to identify young populations are summarised in Table's 1 of \citet{lenkic2023} and \citet{Jones2023_NGC346}.
The area in the CMD where YSOs reside may also be inhabited by any contaminating background galaxies that remain after the cuts described in Section~\ref{sec:sourcedetection} have been applied, but their morphology is distinct from that of a YSO and are easily removed from the catalogue.

\subsection{Luminosity Functions} % Section 4.3
\label{sec:lfs}
\begin{figure} % Figure 4
    \centering
    \includegraphics[width=0.5\textwidth]{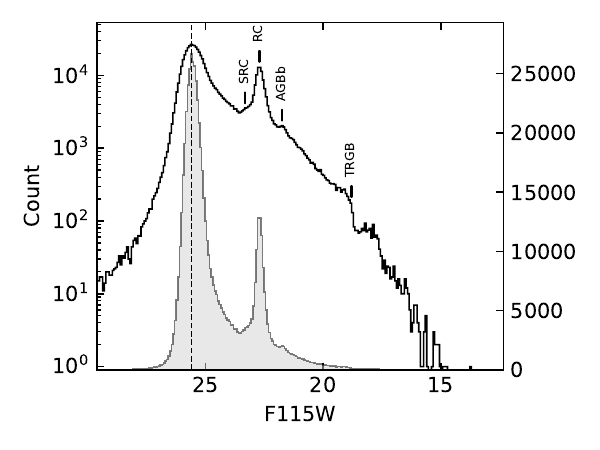}
    \caption{Luminosity function for the F115W filter. Log scaled distribution is drawn in black and linear scaled distributions in filled grey. Bin sizes have been scaled optimally for the number of sources detected. The inferred completeness estimates are denoted with a dashed vertical line and TRGB, AGBb, VRC and SRC locations marked.
    Prior to generating this luminosity function, the F115W catalogue is cropped to F115W$-$F200W$>0.4$  removing the UMS. 
    }
    \label{fig:luminosityfns}
\end{figure}

In \autoref{fig:luminosityfns} we present the luminosity function from our full band merged catalogue in the F115W filter. The logarithmic scaled distribution is plotted in black and its linearly scaled version is plotted in filled gray. The optimal bin width is calculated for the number of sources present in each case using Knuth's Rule \citep{Knuth2006}.  We crudely estimate the completeness of our sample in each JWST filter by identifying the location of the turnover of the luminosity functions at the faint end of the magnitude distribution for each band, marked in the figure with a dashed line. 

We measure the TRGB for every filter in our catalogue sensitive enough to detect it. The TRGB is an important feature from an astrophysical perspective as it is a well-constrained standard candle, depending only weakly on age and metallicity~\citep{Cioni2000, McQuinn2019, Freedman2020}. It represents the final stage of evolution up the RGB before the onset of helium core burning, moving the star off this branch and onto either the HB, RC or VRC depending on the age, metallicity and mass of the star. At this upper boundary, RGB stars have a hydrogen-burning shell supported by an electron-degenerate helium core. This prominent feature is used to separate the RGB stars from the TP-AGB stars which are typically ($\sim$90$\%$) brighter than the TRGB~\citep{bib:Boyer2015a}. 
Due to molecular emissions in the photosphere of the RGB stars, we expect the TRGB to not be flat but to slope upwards in the IR~\citep[e.g.,][]{McQuinn2019, Cerny2020, Durbin2020}.
This results in a less steep drop in a luminosity function than would be expected in the optical. This spread in the measured JWST TRGB magnitudes and corresponding uncertainty requires careful empirical calibration before use as a standard candle. The TRGB positions measured here are only estimations to provide context for identifying features and stellar populations in the JWST CMDs, a full analysis of the TRGB position and its slope is outwith the scope of the study.

To measure the position of the TRGB we first remove the blue UMS sources with F115W$-$F200W$<0.4$, then we randomly sample a subset of magnitudes from the catalogue and smooth the distribution with a kernel density estimate. The first-order derivative of the smoothed distribution is calculated with a Savitizky-Golay filter and the TRGB is located at the point of steepest decline, seen in the differential as a sharp trough. This process is repeated thousands of times with different samples of the catalogue and the results averaged to determine the TRGB. We quote the error on this measurement from the standard deviation of the calculated TRGB values for multiple iterations of the algorithm which accounts for the distribution of photometric errors in a given magnitude bin \citep[see e.g.,][]{bib:Hirschauer2020, Jones2023_M32}.

The SED of an RGB star steadily falls with increasing wavelength in the IR and will become undetectable towards the mid-IR when an IR excess from circumstellar dust may also contribute to the flux emission from AGB and RSG stars which may cause the most extreme of these to drop below the TRGB.  Furthermore, at these wavelengths, other objects (e.g., YSOs) with cold dust emission may contribute substantially to the integrated flux \citep{Jones2015b}. This makes a clear identification of the TRGB challenging in the MIRI data. We identify the TRGB in MIRI up to and including F1000W.
\autoref{tab:sensitivity} lists the number of sources detected at each wavelength, the corresponding faint and bright source limits for that filter, and where appropriate, the magnitude of the TRGB.

In F115W we are roughly complete to $\mathrm{F115W}=25.57$, around $\sim$2.8 magnitudes below the RC population visible at $\mathrm{F115W}\sim22.1$ mag. This feature is not symmetrical, a shoulder is present towards fainter magnitudes at F200W=22.5, which is formed by the flume of the SRC described in \autoref{sec:vrc}.

The luminosity function also shows the prominent AGBb \citep{Ferraro1999} at $\mathrm{F115W}=21.75$ and the TRGB at $\mathrm{F115W}=18.79$. 
The prominent nature of the position of the TRGB will allow us to constrain the distance to NGC 6822 and improve the understanding of the sloped relation visible in the IR. Above this, we begin to see the AGB form a shoulder to $\mathrm{F115W}=18.0$. Saturation affects the completeness of our catalogue at around $\mathrm{F115W}=15.3$. This low saturation limit will cause our sample to be less sensitive to the brightest sources detected in \jhk{} data in previous surveys.

\begin{table} % Table 2
    \centering
        \caption{Source count per filter and sensitivity limits estimated by locating the turnover at the lower end of the luminosity functions, and upper limits placed around the brightest collection of sources. The TRGB is calculated by finding the point of steepest decline in the luminosity function. All magnitudes are given in the Vega system.}
    \begin{tabular}{ccccc}
        \hline
        \hline
         Filter & Source & Completeness & Bright Limit & TRGB (error)\\
                & Count  &   [VegaMag] & [VegaMag] & [VegaMag]\\
         \hline
         F115W &  792424 & 25.60 & 15.3 & 18.79(04) \\
         F200W &  539377 & 24.58 & 15.0 & 17.51(03) \\
         F356W &  184141 & 23.21 & 14.5 & 17.49(06) \\
         F444W &  155327 & 23.09 & 14.0 & 17.65(12) \\
         F770W &  97817  & 20.43 & 12.5 & 17.43(19) \\
         F1000W & 7098   & 19.41 & 11.5 & 17.40(13) \\
         F1500W & 1311   & 16.80 & 10.0 & - \\
         F2100W & 794    & 16.86 & 9.0 & - \\
         \hline
    \end{tabular}
    \label{tab:sensitivity}
\end{table}

\subsection{Parallel Field Photometry} % Section 4.4
\label{sec:parallelphot}

\begin{figure*}
    \centering
    \includegraphics[width=0.45\textwidth]{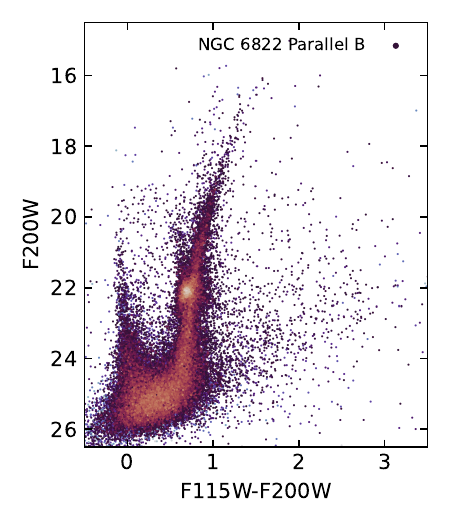}
    \includegraphics[width=0.45\textwidth]{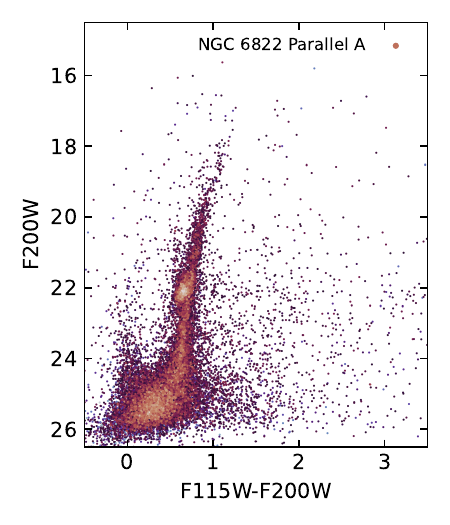}
    \caption{Colour-Magnitude Diagrams $\mathrm{F115W}-\mathrm{F200W}$ vs.\ $\mathrm{F200W}$ in Hess format for the parallel imaging fields. Parallel $\mathrm{B}$ (left) is the NIRCam module B, appearing closest to the centre of NGC~6822, containing the star-forming region Hubble~$\mathrm{IV}$. Parallel $\mathrm{A}$ (right) is the NIRCam module A, the furthest field away from the centre of the galaxy. }
    \label{fig:parallelcmds}
\end{figure*}

The spatial coverage of the two parallel fields is shown in \autoref{fig:FoVs}. NIRCam parallel $\mathrm{B}$ contains the star-forming region Hubble $\mathrm{IV}$. NIRCam parallel $\mathrm{A}$ contains no known areas of active star formation but is contaminated by the PSF spikes from a bright off-field foreground star, which may affect the quality of the photometric extractions for point sources located within these spikes.   

We conduct photometry for the two NIRCam parallel imaging fields and plot the $\mathrm{F115W}-\mathrm{F200W}$\ vs.\ $\mathrm{F200W}$ CMD in \autoref{fig:parallelcmds}. For consistency the \textsc{starbugii} photometric parameters listed in appendix \autoref{tab:sb_params} used for the main field of NGC 6822 are also used for the photometric extraction in these parallel fields. 
We also match our parallel catalogues to \emph{Gaia}~DR3 and remove any foreground contaminant's which meet the criteria given in Section~\ref{sec:FGDcontamination}, prior to generating any CMDs and extinction correct the catalogue as described in Section~\ref{sec:corrections}.
Both parallel fields $\mathrm{A}$ and $\mathrm{B}$ have source detections approximately 0.5 magnitudes deeper than the main field, likely due to the decreased crowding and hence a lower surface brightness in these regions. 

The NIRCam parallel $\mathrm{B}$ field contains young UMS stars seen in \autoref{fig:parallelcmds} as a tight track on the left of the CMD. The RGB turns off and extends upwards to the red, but the TRGB itself is not well defined. The RC sits prominently out of the RGB to the left from which the intermediate-age VRC track extends vertically from it. A faint track of RSG stars splits from the RGB at F200W$\sim$19.5 and is well separated from a small collection of AGB stars above the TRGB calculated in \autoref{tab:sensitivity}.

NIRCam parallel $\mathrm{A}$ does not contain many UMS stars beyond a short scattered track above the MSTO. Similarly, the VRC stars are only marginally present and no obvious RSG sequence is seen. The relative lack of very young stars is due to there being no active star-forming region in the FoV.

\section{Conclusion and Summary} % Section 5.
\label{sec:summary}

We observed the central stellar bar of NGC~6822, utilizing the spatial resolution and sensitivity afforded by JWST NIRCam and MIRI to characterise the IR stellar populations of this isolated metal-poor dwarf galaxy. 
Our observations were designed to image the most dust-enshrouded stars from evolved TP-AGBs to young pre-main-sequence stars and YSOs embedded within the super star cluster candidate Spitzer~I.

\begin{itemize}

    \item We produce JWST photometric catalogues of NGC 6822 with {\sc starbugii} and estimate the point-source sensitivities for these observations which are in the range 25.57 -- 14.0 mag for the NIRCam filters, and 19.59 -- 9.0 mag for the MIRI filters. 

    \item We estimate positions of the TRGB in the NIRCam and MIRI filters up to and including F1000W. The rapidly falling SED of RGB stars in IR wavelengths prevents this feature from being detected  in our F1500W and F2100W MIRI data.
    
    \item We show CMDs of varying colour combinations, using JWST equivalents of UKIRT \jhk{} and {\em Spitzer} filter combinations from previous surveys of NGC~6822 and the Magellanic Clouds to guide the placement of source classifications when pairing NIRCam and MIRI data. We identify several key populations in the NGC 6822 field. 
    
    \item We observe the UMS, RC and VRC populations as well as the elusive and short-lived AGBb phase and demonstrate the position these inhabit in several JWST CMDs.

    \item We detect a population of carbon- and oxygen-rich AGB stars and match them to independently classified IR catalogues to determine the boundary between them. These populations are distinct from one another and separate from a younger RSG population which is also detected.

    \item We show a complex and ongoing star formation history is present in the central bar of NGC 6822 by the detection of old, young and intermediate age stars.

    \item Using parallel near-IR imaging of fields outside the central stellar bar of NGC 6822, we show that the young populations are absent or heavily reduced in number. 

\end{itemize}

%%% **************************************************************************************************** %%%

\

\noindent {\it Facilities:} {\em JWST} (NIRCam \& MIRI) - James Webb Space Telescope.

\noindent {\it Software:} {jhat \citep{jhat}, image1overf.py \citep{1fcor}, astropy \citep{Astropy2013}, {\sc starbug ii} \citep{Starbug2} , and {\sc topcat} \citep{Taylor2005}.}

%%%%%%%%%%%%%%%%%%%%%%%%%%%%%%%%%%%%%%%%%%%%%%%%%%%%%%%%%%%%%%%%%%%%%%%%%%
\section*{Acknowledgements}

The authors are grateful to the anonymous referee who helped them in improving this paper with constructive suggestions. This work is based on observations made with the NASA/ESA/CSA James Webb Space Telescope. The data were obtained from the Mikulski Archive for Space Telescopes at the Space Telescope Science Institute, which is operated by the Association of Universities for Research in Astronomy, Inc., under NASA contract NAS 5-03127 for JWST. These observations are associated with program \#1234.
CN acknowledge the support of an STFC studentship (2645535).
OCJ has received funding from an STFC Webb fellowship.
MM and NH acknowledge support through NASA/JWST grant 80NSSC22K0025 and MM and LL acknowledge support from the NSF through grant 2054178.
MM, NH, and LL acknowledge that a portion of their research was carried out at the Jet Propulsion Laboratory, California Institute of Technology, under a contract with the National Aeronautics and Space Administration (80NM0018D0004).
PJK acknowledge support from Science Foundation Ireland/Irish Research Council Pathway programme under Grant Number 21/PATH-S/9360. AMNF is grateful for support from the UK STFC via grant ST/Y001281/1.
ASH is supported in part by an STScI Postdoctoral Fellowship.
%

%%%%%%%%%%%%%%%%%%%%%%%%%%%%%%%%%%%%%%%%%%%%%%%%%%%%%%%%%%%%%%%%%%%%%%%%%%
\section*{Data availability}

The data used in this study may be obtained from the Mikulski Archive for Space Telescopes (MAST; \href{https://mast.stsci.edu/}{https://mast.stsci.edu/}) and are associated with program \#1234.

\clearpage

%%%%%%%%%%%%%%%%%%%%%%%%%%%%%%%%%%%%%%%%%%%%%%%%%%

%%%%%%%%%%%%%%%%%%%%%%%%%%%%%%%%%%%%%%%%%%%%%%%%%%%%%%%%%%%%%%%%%%%%%%%%%%

%%%%%%%%%%%%%%%%%%%%%%%%%%%%%%%%%%%%%%%%%%%%%%%%%%


%\input{journaldefs} 
\bibliographystyle{mnras}

\bibliography{main} % your references Yourfile.bib

%\clearpage
\appendix
\section{Observational and Photometric Tables}
\begin{table*}
	{\centering
     \caption{Observing Parameters for the NIRCam Prime and Parallel Imaging.}
	\begin{tabular}{ccccccc} % four columns, alignment for each
		\hline
        \hline
        Filter &Field &Readout Pattern &Groups/Int. &Int./Exp. &Dithers &Total Exp.\ Time \\
        & & & & & & [sec] \\
		\hline
        F115W & NIRCam Prime & \textsc{bright2} & 7 & 1 & 12 & 1803.777 \\
        F200W & NIRCam Prime & \textsc{bright2} & 7 & 1 & 12 & 1803.777 \\
        F356W & NIRCam Prime & \textsc{bright2} & 7 & 1 & 12 & 1803.777 \\
        F444W & NIRCam Prime & \textsc{bright2} & 7 & 1 & 12 & 1803.777 \\
        \hline
        F140M & NIRCam Parallel & \textsc{shallow4} & 8 & 1 & 4 & 1674.936 \\
        F335M & NIRCam Parallel & \textsc{shallow4} & 8 & 1 & 4 & 1674.936 \\
        F115W & NIRCam Parallel & \textsc{shallow2} & 3 & 1 & 4 & 515.365 \\
        F150W & NIRCam Parallel & \textsc{shallow2} & 4 & 1 & 4 & 730.100 \\
        F200W & NIRCam Parallel & \textsc{shallow4} & 4 & 1 & 4 & 815.995 \\
        F277W & NIRCam Parallel & \textsc{shallow2} & 4 & 1 & 4 & 730.100 \\
        F356W & NIRCam Parallel & \textsc{shallow2} & 3 & 1 & 4 & 515.365 \\
        F444W & NIRCam Parallel & \textsc{shallow4} & 4 & 1 & 4 & 815.995 \\
		\hline
	\end{tabular}
    \label{tab:obs_params_NIRCam}
    }
\end{table*}

\begin{table*}
	\centering
     \caption{Observing Parameters for the MIRI Prime and Parallel Imaging.}
	\begin{tabular}{ccccccc} % four columns, alignment for each
		\hline
        \hline
        Filter &Field &Readout Pattern &Groups/Int. &Int./Exp. &Dithers &Total Exp.\ Time \\
        & & & & & & [sec] \\
		\hline
        F770W & MIRI Prime & \textsc{fastr1} & 52 & 1 & 4 & 577.208 \\
        F1000W & MIRI Prime & \textsc{fastr1} & 36 & 2 & 4 & 810.312 \\
        F1500W & MIRI Prime & \textsc{fastr1} & 15 & 8 & 4 & 1409.72 \\
        F2100W & MIRI Prime & \textsc{fastr1} & 20 & 9 & 4 & 2086.83 \\
        \hline
        F1000W & MIRI Parallel & \textsc{slowr1} & 5 & 1 & 12 & 1433.395 \\
        F1500W & MIRI Parallel & \textsc{slowr1} & 5 & 1 & 12 & 1433.395 \\
		\hline
	\end{tabular}
    \label{tab:obs_params_MIRI}
\end{table*}

\begin{table*}
    \centering
    \caption{{\sc starbugii} parameters used for all photometry.}
    \begin{tabular}{l|cccccccc}
    \hline
    \hline
        Parameter &F115W & F200W & F356W & F444W & F770W & F1000W & F1500W & F2100W\\
        \hline
        SIGSRC       & 5.0 & 5.0 & 5.0 & 5.0            &5.0&5.0&4.0&3.0 \\ 
        SIGSKY       & 2.0 & 2.0 & 2.0 & 2.0            &1.8&1.8&1.8&1.8 \\ 
        RICKER\_R    & 1.0 & 1.0 & 1.0 & 1.0            &2.0&2.0&5.0&5.0 \\ 
        SHARP\_LO    & 0.4 & 0.4 & 0.55& 0.4            &0.4&0.4&0.225&0.2 \\ 
        SHARP\_HI    & 1.1 & 0.9 & 0.9 & 0.85           &0.76&0.9&0.53&0.84 \\ 
        ROUND\_LO/HI & $\pm1.5$ & $\pm1$ & $\pm1$ & $\pm1$  &$\pm0.6$&$\pm1.0$&$\pm0.6$&$\pm0.5$ \\ 
        \hline
        ENCENRGY     & 0.6 & 0.6 & 0.6 & 0.6            &0.6 &0.6 &0.6 &0.6 \\
        SKY\_RIN     & 3.0 & 3.0 & 3.0 & 3.0            &4.0&4.0&5.0&6.0 \\ 
        SKY\_ROUT    & 4.5 & 4.5 & 4.5 & 4.5            &5.5&5.5&6.5&7.5 \\ 
        BOX\_SIZE    & 15 & 15 & 5 & 5                  &2&5&5&8 \\ 
        CRIT\_SEP    & 5 & 6 & 8 & 8                    &8&8&8&8 \\ 
        \hline
        MATCH\_THRESH & 0.1 & 0.1 & 0.1 & 0.1           &0.15&0.2&0.2&0.2 \\ 
        NEXP\_THRESH    & 2 & 2 & 2 & 2                 &3&3&3&3 \\ 
        \hline
    \end{tabular}
    \label{tab:sb_params}
\end{table*}
%%%%%%%%%%%%%%%%%%%%%%%%%%%%%%%%%%%%%%%%%%%%%%%%%%

%%%%%%%%%%%%%%%%% APPENDICES %%%%%%%%%%%%%%%%%%%%%

%\appendix


% Appendix appears after the list of references.

%%%%%%%%%%%%%%%%%%%%%%%%%%%%%%%%%%%%%%%%%%%%%%%%%%

% Don't change these lines
\bsp	% typesetting comment
\label{lastpage}
\end{document}